# Automated Region Masking Of Latent Overlapped Fingerprints

Tejas K, Swathi C, Aravind Kumar D, Rajesh Muthu, *Senior Member, IEEE*
Department of Electronics and Communication Engineering
VIT University
Vellore, India 632014
Email: tejastk.reddy@gmail.com, mrajeshkumar@vit.ac.in

*Abstract*—**Fingerprints have grown to be the most robust and efficient means of biometric identification. Latent fingerprints are commonly found at crime scenes. They are also of the overlapped kind making it harder for identification and thus the separation of overlapped fingerprints has been a conundrum to surpass. The usage of dedicated software has resulted in a manual approach to region masking of the two given overlapped fingerprints. The region masks are then further used to separate the fingerprints. This requires the user's physical concentration to acquire the separate region masks, which are found to be time-consuming. This paper proposes a novel algorithm that is fully automated in its approach to region masking the overlapped fingerprint image. The algorithm recognizes a unique approach of using blurring, erosion and dilation in order to attain the desired automated region masks. The experiments conducted visually demonstrate the effectiveness of the algorithm.**

I. INTRODUCTION

Fingerprints are unique to each individual and thus are the primary means of identification recognized at the global level. Fingerprints consist of ridges in patterns unique to each individual. Ridges vary and exist as two major patterns ridge endings and ridge bifurcations. These are called minutiae and used in recognition and identification.

Latent overlapped fingerprints are most commonly found at crime scenes as shown in Fig.1. Latent fingerprints themselves provide various difficulties as a means of identification. Significant work prevails in separation of overlapped fingerprints. In recent research, technology has made advancements in overlapped fingerprint separation for further recognition [1], [2], [3] and [4]. The work done by Chen et al [4] is an advanced and distinct algorithm. This was further developed by Jie Zhou [2], who proposed a better algorithm for constrained relaxation labeling and Jianjiang Feng [3], who improved the efficiency of program to estimate the initial orientation field, added to the standards of the algorithm proposed in [4]. Recent technologies have made the algorithm work faster with minimal manual markup.

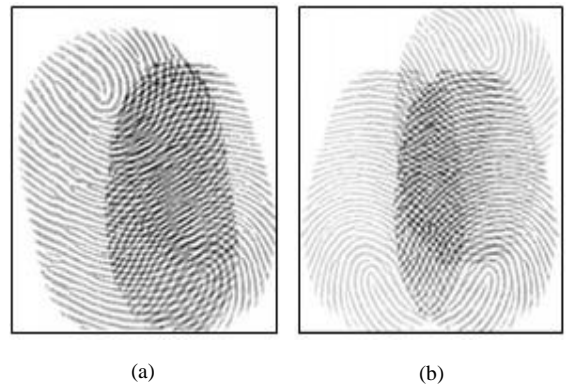

Fig.1. Latent Overlapped Fingerprints (a) Two Fingerprints Overlapped (b) Three Fingerprints overlapped

The algorithm proposed by [4] involve five unique steps, namely manual region masking, initial orientation field estimation, constrained relaxation labeling using local Fourier analysis, error correction and image enhancement using Gabor filters to separate the component fingerprints.

The primary step in the separation is region masking. Region masking is the process of separation of an overlapped fingerprint into background and foreground regions, the foreground region is further sub-divided into overlapped region and non-overlapped regions of each individual component fingerprints. The region mask helps in segregating different regions of overlapped fingerprint, which are further used to estimate the initial orientation field. The research made so far within overlapped fingerprint separation employs the use of a manual approach to carry out region masking. The referred papers have said that one of the limitations in their paper is the manual approach to region masking which is a tedious and time consuming task.

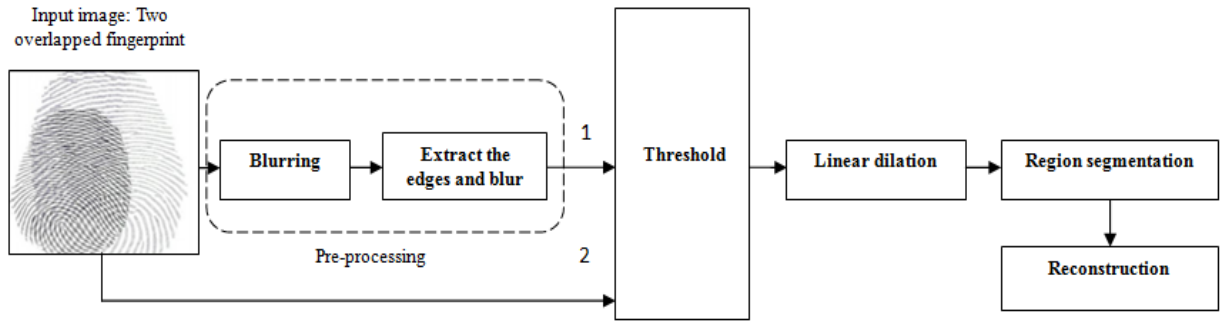

Fig.2. Block diagram of proposed region segmentation algorithm; (1) Edge extracted and blurred image; (2) Input gray scale image.

This paper proposes a novel algorithm to fully automated method not only hopes to reduce time, but also produce more efficient results. This eliminates the physical exertion in obtaining the two fingerprint regions and also the time spent is significantly reduced. This would also reduce the labor of skilled fingerprint examiners. The unique approach uses basic commands in a sequential manner so as to arrive at a fully automated approach.

## II. PRE-PROCESSING

The initial step involves processing the given overlapped fingerprint image for future steps. Any inconveniences that may occur later in the algorithm are considered and precautions are implemented in this portion of the algorithm. Pre-processing consists of the following steps:
1. Multiple rows and columns are padded to the original image so as to resize it to have a larger working space. This is so that no inconvenience is encountered later in the dilation process.
2. The resultant image is now blurred with the help of an averaging filter. Experimentally, an averaging filter of size 15-20 rows and columns respectively, yielded better results. The blurring is done in such a way that the central part of the correlation is returned.
3. Now the above-formed image is subtracted from a plain white image of the same size in order to extract the edges from the blurred image. This image obtained after subtraction is further blurred with the help of an averaging filter. The blurring is done in such a way that the parts of the correlation that are computed without zero-padded edges are returned.
4. This image is multiplied by a constant. Experimentally, on trial and error process it was observed that multiplying with a constant value of 1.5 to 5.5 yielded the best results in the further part of the program. This constant value is directly proportional to the contrast of the fingerprints in the input image. This is done so that the correlation is better viewed.

## III. THRESHOLD

The second part of the algorithm now consists of removing the overlapped region within the fingerprint. Observing the difference in intensities of the image formed in step 3 of preprocessing, threshold the image to obtain the desired portion. Threshold is later applied on the input image to get the accurate region of interest, which segregates the background and foreground regions of input image.
1. It is observed from the image obtained after step 4 of pre-processing, that a certain part of the image has relatively higher intensity as compared to other parts. This intensified part of the image represents the overlapped region of input image. This is clearly illustrated in Fig.3. (c). Now suitable threshold is applied to extract the overlapped region.
2. Our concentration now shifts to separating the entire overlapped fingerprint image from the background. Obtaining the entire region will prove to be helpful in the reconstruction of the overlapped image for validation. This can be done by applying a high threshold value to the image [5]. We now 'area open' the binary image produced after applying threshold to the input overlapped fingerprint. This is done to remove the noise and to get better results in further part of the algorithm. The resultant image consists of the boundary of initially given overlapped fingerprints as shown in Fig.3. (e).

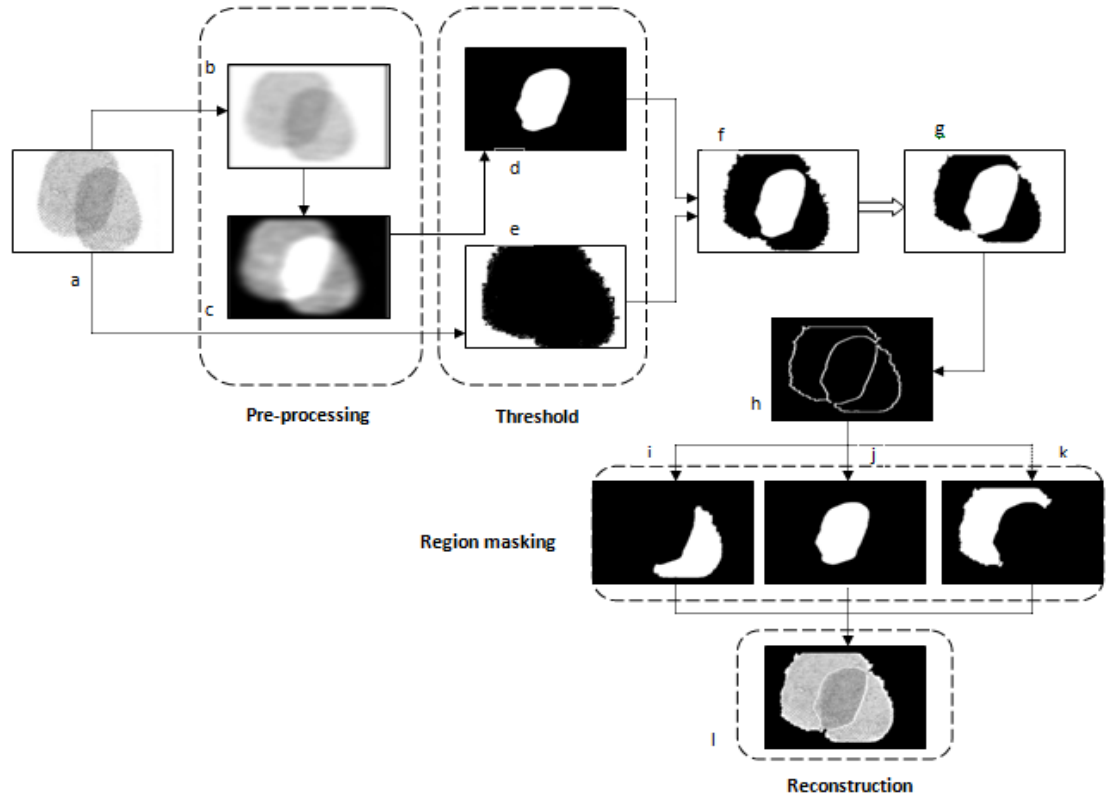

Fig.3. Flowchart of the proposed region masking algorithm. Here the input image (a) is padded with additional rows and columns  (b) blurred input image  (c) edge extracted and blurred image  (d) after applying threshold on image c  (e) applying threshold on input image  (f) addition of images  d and  e  (g) linear dilation  (h) Sobel edge extraction  (i) region mask of first component fingerprint  (j) overlapped region  (k) region mask of second component fingerprint  (l) reconstructed image.

## IV. AUTOMATIC REGION MASKING

The success of the algorithm lies in accurately region masking the non-overlapped regions of each individual components and the overlapped region common to all component fingerprints. This is achieved from the following steps.

1. The images formed in step 1 and step 2 of threshold application, are now added to each other. After addition, we find that there is a gap between the boundary and overlapped region of the fingerprint. This gap is caused due to the blurring effect in the pre-processing work. This gap resists us in forming an efficient region mask. Thus, in order to join them, we linearly dilate the image in a particular angle suited by the fingerprint. The size of the structuring element used for linear dilation varies according to the input image.

2. Once the overlapped region touches the boundary, extract the edges of the image using Sobel edge algorithm. Now, we extract all the connected components from the binary image one by one in different figures, checking on the attribute perimeter.

3. After extracting the edges, always dilate the edges with 3 x 3 structuring elements to strongly connect the boundaries and to eliminate small gaps. Now, we fill the connected boundaries with holes.

4. The resultant region masks show an error as compared to the original image; this error is due to the linear dilation, which was performed to eliminate the gap between overlapped region and the boundary. Thus, in order to remove this error, we perform the     inverse of it. Invert all the binary images in terms of 0's and 1's. Now linearly dilate the each individual figure image with the same structuring element which was used in above linear dilation. This nullifies the effect of initial linear dilation.

5. Here we obtain different figures, among which one is overlapped region and others are non-overlapped regions of various fingerprints. These region masked binary figures are further used for estimation of initial orientation field, which is a vital role in the separation of latent overlapped fingerprints.

## V. EXPERIMENTAL RESULTS

The goal of the algorithm is to region mask the given overlapped fingerprint automatically into foreground and background regions. The foreground region is further region segmented into overlapped region and remaining two or more individual component regions, so as to reduce the need for the preexisting manual method of marking the fingerprints.

We utilize the official FVC data sets [5] and [6] to validate the region segmentation algorithm. Multiple fingerprints were used from the FVC 2004 and FVC 2006 databases to stimulate the required overlapped database. The individual fingerprints in these datasets are fused to procure the required overlapped fingerprints. A few fingerprints were rotated to certain angles as desired to meet the assumptions. A few genuine resulted overlapped fingerprints are shown in Fig.4, where the first two rows of images are of two overlapped fingerprints whereas the last rows of images show the three overlapped fingerprints.

We have visually demonstrated the effectiveness of above-proposed algorithm for two component fingerprints overlapped in Fig.4, where the first column represents the input overlapped fingerprint, the second and third columns illustrates the non-overlapped regions of first and second fingerprint components respectively. The fourth column represents the overlapped region common to both first and second fingerprint components. The reconstruction is done in the fifth column where various segments are colored for better understanding, to visually compare the resultant image with the input image.

The proposed algorithm works efficiently for three overlapped fingerprints also. This is clearly illustrated with two examples in Fig.5. The non-overlapped regions of three component fingerprints were segmented along with the overlapped region common to all three fingerprints. The overlapped region extracted might consist of sub-overlapped regions, which can be separated by implementing the above algorithm, considering this to be the input image.

The above-demonstrated algorithm not only is successful for relatively good images, but it also proves right for images with very low brightness, images with low contrast, images with very high contrast and images with several types of noises like Gaussian noise, Salt and Pepper noise, Poisson noise and Speckle noise. This is effectively illustrated in Fig.6. Speckle noise and Poisson noise distort the background to certain extent. Therefore, low quality images suffer the risk of losing information in the non overlapped regions of component fingerprints. These noises don't affect the result produces from good quality images.

Manual region masking has to be done in the lab along with other efficient software's which delays the result period. This problem is resolved by automatic region masking algorithm in MATLAB software. Several images were tested and the time of execution was recorded by MATLAB software. It roughly takes three seconds to region mask a given latent overlapped fingerprint image, which is a very short span of time as compared to the manual marking procedure. This proves the speed and efficiency of the algorithm.

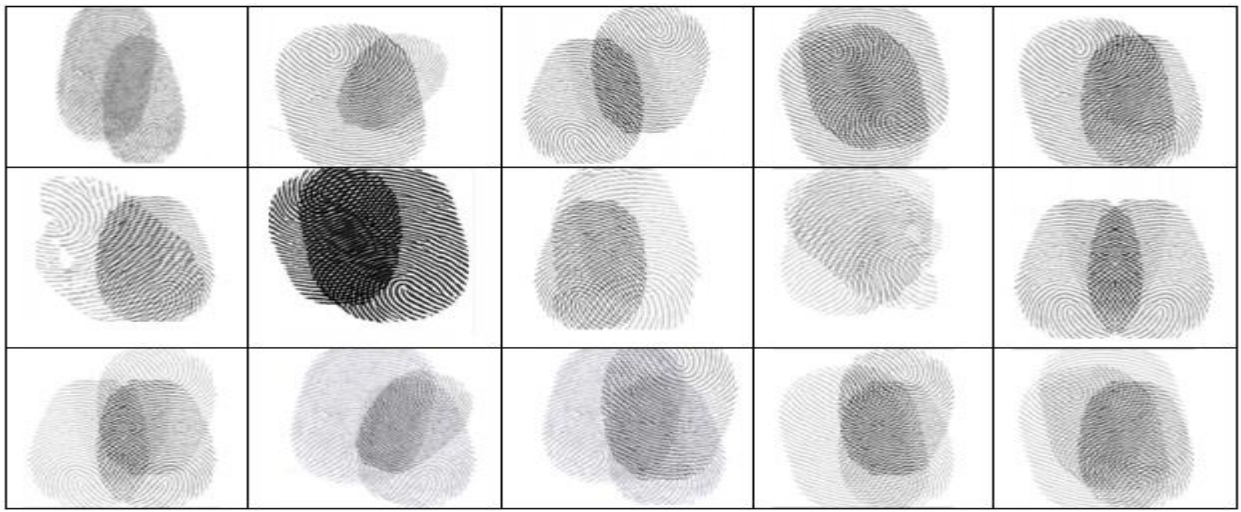

Fig.4. First and second column represent two overlapped fingerprints, third column represents three overlapped fingerprint database.

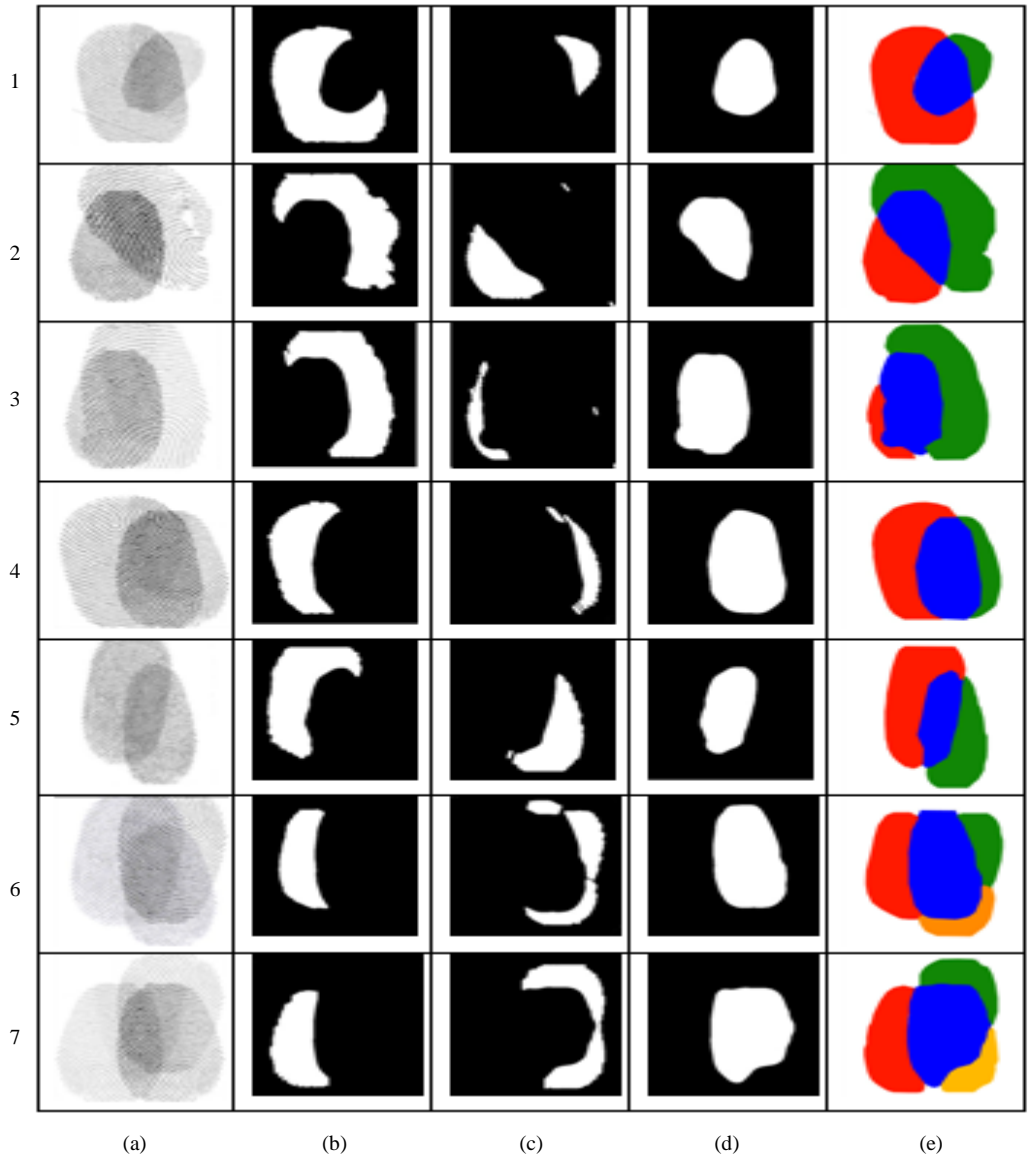

Fig.5. Results of processing two and three overlapped fingerprints by the above proposed algorithm. Rows 1-5 corresponds to two fingerprints overlapped images and rows 6, 7 represents three fingerprints overlapped image. (a) Input image; (b) First fingerprint non-overlapped fingerprint; (c) Remaining fingerprints non-overlapped fingerprints; (d) Overlapped region; (e) Reconstructed region masked image

## VI. CONCLUSION AND FUTURE WORK

Separation of an overlapped fingerprint image is a challenging task. The primary step in the separation is region masking the overlapped fingerprint. The cumbersome task of manually marking the overlapped regions is thus eliminated by this novel algorithm. This algorithm combined with the existing technology could give faster fingerprint matching results. The algorithm is focused on blurring and threshold the given image to obtain the overlapped region as well as the remaining component regions. These region masked components are further used to estimate the initial orientation field.

Thus, this method of automated region masking reduces the time consumed and manual labor required to carry out this task as indicated by the results.

As a part of future work, accuracy can be improved on region masking latent overlapped fingerprints with highly distorted backgrounds or dark backgrounds.

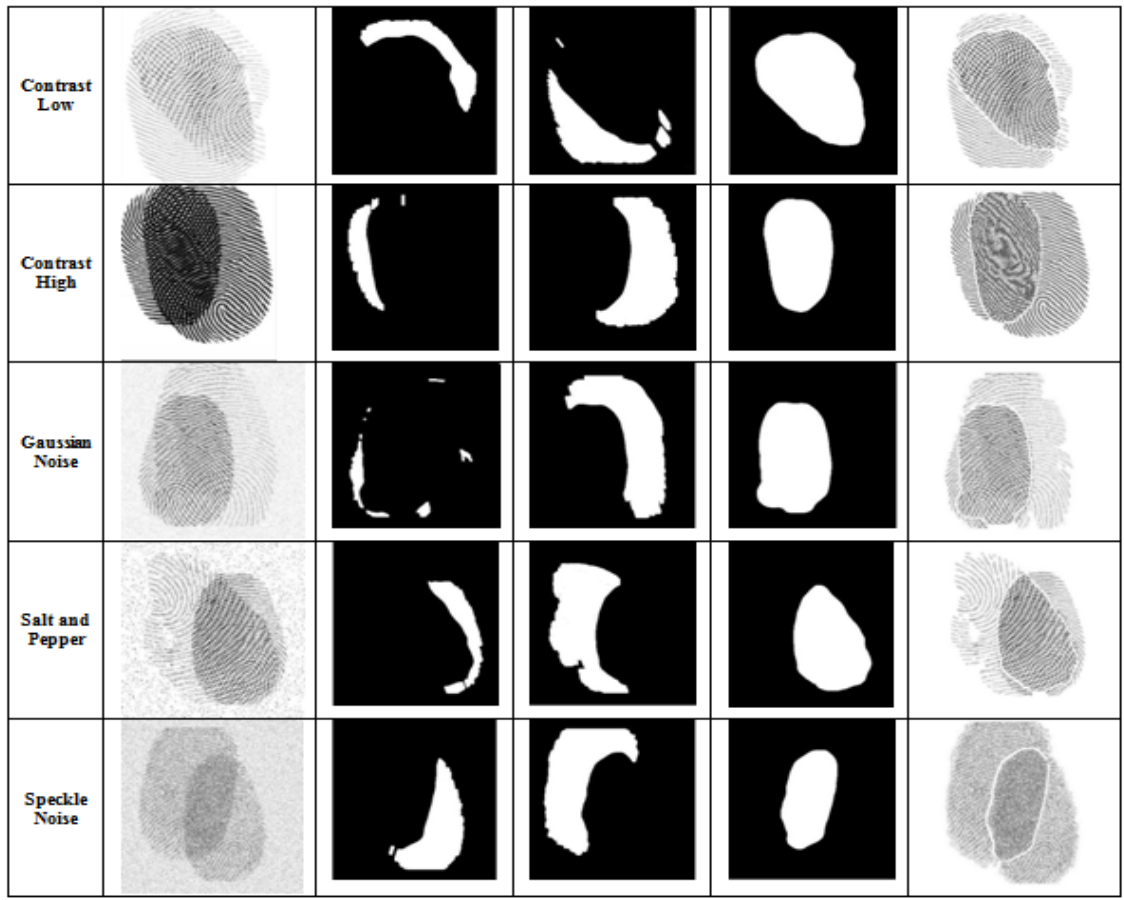

Fig.6. Results of input image subjected to various conditions.